\documentclass[prl,showpacs,superscriptaddress,twocolumn]{revtex4}

\usepackage{graphicx}

\begin{document}

\title{Information theoretic approach to single-particle 
and two-particle interference in multi-path interferometers}

\author{Dagomir Kaszlikowski}
\affiliation{Department of Physics, National University of Singapore, 
Singapore 117\,542, Singapore}

\author{Leong Chuan Kwek}
\affiliation{National Institute of Education, 
Nanyang Technological University, 
Singapore 637\,616, Singapore}
\affiliation{Department of Physics, National University of Singapore, 
Singapore 117\,542, Singapore}

\author{Marek \.Zukowski}
\affiliation{Instytut Fizyki Teoretycznej i Astrofizyki,
Uniwersytet Gda\'nski, 80-952 Gda\'nsk, Poland}

\author{Berthold-Georg Englert}
\affiliation{Department of Physics, National University of Singapore, 
Singapore 117\,542, Singapore}

\date{18 February 2003}   

\begin{abstract}
We propose entropic measures for the strength of single-particle 
and two-particle interference in interferometric experiments where 
each particle of a pair traverses a multi-path interferometer.
Optimal single-particle interference excludes any two-particle 
interference, and vice versa.
We report an inequality that states the compromises allowed 
by quantum mechanics in intermediate situations, and identify a class 
of two-particle states for which the upper bound is reached.
Our approach is applicable to symmetric two-partite systems 
of any finite dimension. 
\end{abstract}

\pacs{03.67.-a, 07.60.Ly}

\maketitle

Interference effects of two kinds can be observed in quantum processes that 
involve pairs of particles.
There are the usual single-particle interference 
fringes and there are phase-dependent coincidence probabilities that constitute
two-particle interferences (see, e.g., \cite{Horne+1:85,Greenberger+2:93}). 
As Horne and Zeilinger noted already in 1985 \cite{Horne+1:85}, there is a
certain complementarity between the two types of interference:
Optimal interference of one kind excludes any interference of the other.

The various quantitative studies of the possible trade-off between
single-particle and two-particle interference that are on record (see
\cite{Jaeger+2:93,Jaeger+2:95} in particular) are dealing with 
two-path interferometers for the single particles.
It is then possible to measure the strength of the single-particle
interference by the familiar Michelson fringe visibility ---the difference 
of maximal and minimal probabilities in one output divided by their sum---
and introduce a matching measure for the two-particle interference.

There is, however, no unique analog for multi-path interferometers, even for
single-particle interference \cite{Weihs+3:96,Zukowski+2:97}.
D\"urr has recently compiled a list of desirable properties of any multi-path 
generalization of Michelson's two-path visibility \cite{Durr:01}, but quite a
few of equally plausible definitions meet the criteria \cite{Englert+2:03},
and the extension to two-particle interference is largely unexplored
territory.

In an attempt to close this gap, at least partly, we introduce here an 
information-theoretic definition of the measure of the strength of 
two-particle interference. 
It exploits the mutual information contained in the coincidence probabilities.
The corresponding single-particle measure is, in some sense, 
the fringe-contrast analog of the entropic measure for path knowledge that
was first used by Wootters and \.Zurek \cite{Wootters+1:79} in 1979 in the
context of a two-path interferometer (Young's double slit), and later also by
others (see, e.g., \cite{Mittelstaedt+2:87,Lahti+2:91}).
But entropic measures of this kind have never been popular and have been
criticized occasionally \cite{Jaeger+2:95,Durr:01}.

Yet, we think that they possess certain advantages. 
In particular, there is the immediate benefit that the dimensionality of the
systems is irrelevant (although we assume, for the purposes of this paper, 
that the two systems are of the same dimension). 

Let us close these introductory remarks by mentioning that the
problem studied here is very closely linked with the relations between
single-particle fringe visibility and which-path knowledge in quantitative
studies of wave-particle duality. 
Going back to Einstein's 1905 paper on the photo-electric effect, this
intriguing aspect of quantum mechanics is as old as quantum theory itself. 
Recent studies include 
\cite{Jaeger+2:93,Jaeger+2:95,Englert:96,Englert+1:00,Durr:01,Jakob+1:03} 
where also the earlier literature can be found.

\begin{figure}[!b]
\includegraphics{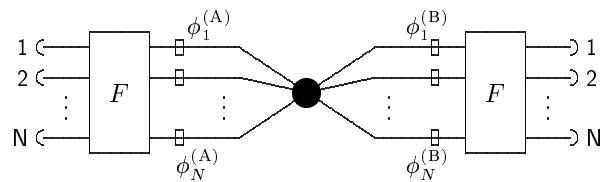}
\caption{\label{figure}%
Sketch of the interferometric setup.
The source at the center emits paired particles. 
One of them propagates to the left, the other to the right, whereby both have
a choice between $N$ paths.
After traversing the respective arrays of phase shifters (phases
$\phi^\mathrm{(A)}_k$ on the left, $\phi^\mathrm{(B)}_l$ on the right)
the particles pass through $N$-port beam splitters that effect a discrete
Fourier transformation ($F$) of the probability amplitudes for the paths, 
and are then detected.}
\end{figure}

Consider an interferometric experiment of the kind that is 
sketched in Fig.~\ref{figure}.
A source emits pairs of entangled particles, one of each pair propagating to 
observer Alice (A) on the left, the other to Bob (B) on the right, $N$ paths
being available on either side. 
Before its registration by the respective detectors, each particle undergoes 
a unitary transformation consisting of the action of phase shifters 
(phases $\phi^\mathrm{(A)}_k$ and $\phi^\mathrm{(B)}_l$, respectively) and 
an unbiased symmetric multiport beam splitter.
The latter effects the unitary discrete Fourier transformation, 
$F$, that is specified by the
matrix elements $F_{mn}=\gamma_{N}^{mn}/\sqrt{N}$ where
$\gamma_{N}=\exp{(2\pi i/N)}$ is the basic $N$-th root of unity.  
Depending on the two-particle state emitted by the source, the resulting
paired detector clicks may exhibit correlations that originate in the path
interference that is tested by the Fourier transformation in conjunction with
the phase shifters. 
It is our objective here to quantify the strength of the interference 
observed in experiments of this kind.  

The experimental data are the $N^2$ coincidence probabilities 
$p^\mathrm{(AB)}_{mn}=%
p^\mathrm{(AB)}_{mn}(\phi^\mathrm{(A)}_1,\dots,\phi^\mathrm{(A)}_N;%
\phi^\mathrm{(B)}_1,\dots,\phi^\mathrm{(B)}_N)$, of unit sum,
between Alice's $m$-th detector and Bob's 
$n$-th detector ($m,n=1,2,\dots,N$).
The relative frequencies of Alice's and Bob's individual detection events are
given by the marginal sums of $p^\mathrm{(AB)}_{mn}$,
\begin{eqnarray}
  \label{eq:1}
p^\mathrm{(A)}_{m}(\phi^\mathrm{(A)}_1,\dots,\phi^\mathrm{(A)}_N) 
&=&\sum_{n=1}^N p^\mathrm{(AB)}_{mn}\,,\nonumber\\
p^\mathrm{(B)}_{n}(\phi^\mathrm{(B)}_1,\dots,\phi^\mathrm{(B)}_N) 
&=&\sum_{m=1}^N p^\mathrm{(AB)}_{mn}
\,.
\end{eqnarray}
In their dependence on the local interferometric phases, $p^\mathrm{(A)}_{m}$
and $p^\mathrm{(B)}_{n}$ contain the pertinent information about the 
phase relations between the $N$ paths on Alice's and Bob's side,
respectively, and information about the $N^2$ path pairs can be extracted 
from the coincidence probabilities $p^\mathrm{(AB)}_{mn}$.  

We have mentioned above that Michelson's standard definition of the fringe
visibility for two-path interferometers has no unique
generalization for multi-path interferometers.
This is so for a simple reason: 
a single number cannot do full justice to the complex pattern displayed by 
$p^\mathrm{(A)}_{m}(\phi^\mathrm{(A)}_1,\dots,\phi^\mathrm{(A)}_N)$
or $p^\mathrm{(B)}_{n}(\phi^\mathrm{(B)}_1,\dots,\phi^\mathrm{(B)}_N)$.
Thus one must make a judicious choice among the various possible definitions
that meet the reasonably obvious criteria on D\"urr's list \cite{Durr:01}.

Here we opt for measuring
the various interference strengths in terms of the respective Shannon
entropies. 
We shall continue to speak of 
``visibility'' although the analogy with the usual two-path visibility is a
bit remote. 
Alice's single-particle visibility $V^\mathrm{(A)}$ is then defined 
in accordance with
\begin{equation}
  \label{eq:2}
V^\mathrm{(A)}=
\max_{\mbox{\scriptsize$\phi^\mathrm{(A)}_1,\dots,\phi^\mathrm{(A)}_N$}}
\bigl[1-H_N\bigl(p^\mathrm{(A)}\bigr)\bigr]\,,
\end{equation}
where
\begin{equation}
  \label{eq:3}
  H_N\bigl(p^\mathrm{(A)}\bigr)
=-\sum_{m=1}^N p^\mathrm{(A)}_{m} \log_N p^\mathrm{(A)}_{m}
\end{equation}
is the normalized Shannon entropy of her marginal probability distribution,
and Bob's  single-particle visibility $V^\mathrm{(B)}$ is defined analogously. 
For convenient normalization, we take the logarithm to base $N$ so that,
by construction, both $H_N\bigl(p^\mathrm{(A)}\bigr)$ and $V^\mathrm{(A)}$ are
nonnegative numbers that do not exceed unity.

The definition of $V^\mathrm{(A)}$ in (\ref{eq:2}) exploits the fact  
that path knowledge after a Fourier transformation corresponds to
interference strength before the transformation \cite{Englert+2:03}.
The largest amount of terminal path knowledge that is potentially available
for optimally set phase shifters is thus a self-suggesting measure for the
interference strength at the input, and the Shannon entropy is one natural 
quantitative measure of knowledge about a probability distribution---the one
we find particularly useful here.
This a priori justification of our definition of $V^\mathrm{(A)}$ is, of
course, no more than an appeal at plausibility; the definition 
is ultimately justified by the consequences that we report below. 

We note that $V^\mathrm{(A)}$ of (\ref{eq:2}) reaches its upper bound of
unity only for certain pure states of Alice's particle, namely those for
which it is possible to adjust the phases $\phi^\mathrm{(A)}_k$ such that the
particle is surely directed to one particular detector, which is to say
that the arriving particle must follow each path with the same probability 
of $1/N$ and that there must be definite relative phases between the paths. 
It is equally important that $V^\mathrm{(A)}$ vanishes if Alice's particle
arrives along one particular path, in which situation there is no room for
relative phases between the paths.
Accordingly, $V^\mathrm{(A)}$ vanishes also if the statistical operator
$\rho^\mathrm{(A)}$ of Alice's particle is a convex sum of pure
states that refer to definite paths. 
In the case of the totally mixed state $\rho^\mathrm{(A)}=1/N$,
we face the extreme situation of utter ignorance in which nothing is 
predictable about the particle's path before or after the Fourier 
transformation.

The single-particle visibility of (\ref{eq:2}) also
meets the indispensable criterion of convexity: 
The visibility of a convex sum of states cannot exceed the convex sum of 
the individual visibilities.
This is an immediate consequence of the concavity of the Shannon entropy.

As a preparation for the definition of the two-particle visibility 
$V^\mathrm{(AB)}$ in (\ref{eq:5}) below, we now consider the interferometric 
experiment from the perspective of information theory. 
Imagine that the source and beam splitters are contained in a black box and
only the detectors are accessible.
We observe paired clicks of the detectors with the coincidence  
probabilities $p^\mathrm{(AB)}_{mn}$ and, on either side, 
individual clicks with the marginal probabilities $p^\mathrm{(A)}_{m}$
and $p^\mathrm{(B)}_{n}$, respectively.
We then wonder about the strength of correlations between clicks of pairs of
detectors at opposite sides of the experiment. 
Obviously, the quantitative measure of two-particle interference should 
vanish if there are no correlations between the clicks for any 
setting of the phase shifters and it should be maximal  
if perfect correlations can be achieved for some setting.

A good candidate for a measure having these properties, and the natural one
to accompany the single-particle visibility of (\ref{eq:2}), 
is the maximal value of the mutual information $I\bigl(p^\mathrm{(AB)}\bigr)$
contained in the $p^\mathrm{(AB)}_{mn}$'s, which is
the relative Shannon entropy between the probability distributions 
$p^\mathrm{(AB)}_{mn}$ and $p^\mathrm{(A)}_{n}p^\mathrm{(B)}_{m}$, that is: 
between the actual coincidence probabilities and the corresponding products 
of the marginal probabilities.

For a particular setting of the phase shifters, we have 
\begin{eqnarray}
  \label{eq:4}
 I\bigl(p^\mathrm{(AB)}\bigr)&=&
H_N\bigl(p^\mathrm{(A)}\bigr)+H_N\bigl(p^\mathrm{(B)}\bigr)
-H_N\bigl(p^\mathrm{(AB)}\bigr)
\nonumber\\
&=&\sum_{m,n=1}^N p^\mathrm{(AB)}_{mn}\log_N
\frac{p^\mathrm{(AB)}_{mn}}{p^\mathrm{(A)}_{n}p^\mathrm{(B)}_{m}}\,.
\end{eqnarray}
If the two distributions happen to be equal,
$p^\mathrm{(AB)}_{mn}=p^\mathrm{(A)}_{n}p^\mathrm{(B)}_{m}$,
then the stochastic variables $n$ and
$m$ are statistically independent, the detector clicks are not correlated,
and the mutual information vanishes accordingly.
Since it is also bounded, $0\leq I\bigl(p^\mathrm{(AB)}\bigr)\leq1$, 
we are invited to define the two-particle visibility as
\begin{equation}
\label{eq:5} 
V^\mathrm{(AB)}=
\max_{\mbox{\scriptsize$\phi^\mathrm{(A)}_k,\phi^\mathrm{(B)}_l$}}
I\bigl(p^\mathrm{(AB)}\bigr)\,.
\end{equation}

Before proceeding we should not fail to mention the alternative proposal by
Jaeger, Horne, and Shimony \cite{Jaeger+2:93}.
In the context of two-path interferometers, they define a two-particle
visibility by a Michelson-type formula for the extremal values of the 
\emph{difference} 
$p^\mathrm{(AB)}_{mn}-p^\mathrm{(A)}_{n}p^\mathrm{(B)}_{m}$,
and it appears that a systematic study of the possible generalization to
multi-path interferometers is lacking.
By contrast, our definition involves (the  logarithm of) the \emph{ratio} 
$p^\mathrm{(AB)}_{mn}/\bigl(p^\mathrm{(A)}_{n}p^\mathrm{(B)}_{m}\bigr)$
and is immediately applicable to multi-path interferometers.

We note that the two-particle visibility (\ref{eq:5}) is maximal, 
$V^\mathrm{(AB)}=1$, if the source emits a maximally entangled state such as
\begin{equation}
  \label{eq:6}
  \rho^\mathrm{(AB)}_\mathrm{max.\,ent.}=\frac{1}{N}\sum_{j,k=1}^N
|jj\rangle\langle{kk}|\,,
\end{equation}
where $|jk\rangle$ is the ket vector for one particle in Alice's $j$-th path
and the other in Bob's $k$-th path.
And if the source emits a product state,
\begin{equation}
  \label{eq:7}
  \rho^\mathrm{(AB)}=\rho^\mathrm{(A)}\rho^\mathrm{(B)}\,,\enskip
\rho^\mathrm{(A)}=\mathrm{tr}_\mathrm{B}\rho^\mathrm{(AB)}\,,\enskip
\rho^\mathrm{(B)}=\mathrm{tr}_\mathrm{A}\rho^\mathrm{(AB)}\,,
\end{equation}
then $p^\mathrm{(AB)}_{mn}=p^\mathrm{(A)}_{n}p^\mathrm{(B)}_{m}$
and $V^\mathrm{(AB)}=0$.
This limit is also reached in the case of the maximally mixed, or chaotic, 
state
\begin{equation}
  \label{eq:8}
   \rho^\mathrm{(AB)}_\mathrm{chaos}=\frac{1}{N^2}\sum_{j,k=1}^N
|jk\rangle\langle{jk}|=\frac{1}{N^2}\,.
\end{equation}
All these features are in complete agreement with what one would expect from
a reasonably defined two-particle visibility.

The stage is now set for presenting the central result:
For correlations generated by quantum mechanics 
(arbitrary joint probabilities $p^\mathrm{(AB)}_{mn}$ 
do not have this property), the inequalities
\begin{equation}
  \label{eq:9}
 V^\mathrm{(A)}+ V^\mathrm{(AB)}\leq1\,,\quad
 V^\mathrm{(B)}+ V^\mathrm{(AB)}\leq1
\label{comp} 
\end{equation}
are obeyed by the single-particle visibilities $V^\mathrm{(A)}$,
$V^\mathrm{(B)}$ and the two-particle visibility $V^\mathrm{(AB)}$,
and there are families of pure two-particle states, 
$\rho^\mathrm{(AB)}_\lambda=|\lambda\rangle\langle\lambda|$ 
with $0\leq\lambda\leq1$, for which the equal signs hold with
each visibility covering the whole range from $0$ to $1$.
In other words, the inequalities can be saturated.

\begin{figure}[!t]
\includegraphics{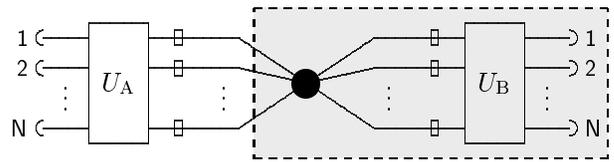}  
\caption{\label{figure2}%
Another view of the experiment. 
It is \emph{as if} Bob prepared Alice's particle on the left in a particular
state by detecting his particle on the right. 
Upon regarding the components inside the shaded area as Bob's effective
source, the setup can be viewed as realizing a certain communication
protocol, and can thus be interpreted and assessed from the point of view of
information theory.}
\end{figure}

In view of the complete symmetry between Alice and Bob, it suffices to prove
one of the inequalities, say the first one.
Consider the experiment of Fig.~\ref{figure2}, where 
we have two observers who can apply arbitrary $U(N)$ transformations,
$U_\mathrm{A}$ and $U_\mathrm{B}$, to
their respective incoming particles, which arrive in some state
$\rho^\mathrm{(AB)}$ that could be pure or mixed. 
Since there is no restriction here to Fourier transformations, the more
general single-particle visibility $\widetilde{V}^\mathrm{(A)}$ bounds 
$V^\mathrm{(A)}$ from above,
\begin{equation}
  \label{eq:10}
 V^\mathrm{(A)}\leq\widetilde{V}^\mathrm{(A)}=
\max_{\mbox{\footnotesize$U_\mathrm{A}$}}
\bigl[1-H_N\bigl(p^\mathrm{(A)}\bigr)\bigr]
=1-S_N\bigl(\rho^\mathrm{(A)}\bigr)\,,
\end{equation}
where $S_N(\rho)=-\mathrm{tr}\bigl\{\rho\log_N\rho\bigr\}$
is the scaled von Neumann entropy of the reduced state $\rho^\mathrm{(A)}$.
The maximum in (\ref{eq:10}) is reached for those $U_\mathrm{A}$ that
diagonalize  $\rho^\mathrm{(A)}$.

To find a corresponding upper bound for the two-particle visibility, we first
note that
\begin{eqnarray}
  \label{eq:12}
  V^\mathrm{(AB)}\leq\widetilde{V}^\mathrm{(AB)}&=&
 \max_{\mbox{\footnotesize$U_\mathrm{A},U_\mathrm{B}$}}
\bigl[H_N\bigl(p^\mathrm{(A)}\bigr)+H_N\bigl(p^\mathrm{(B)}\bigr)
\nonumber\\
&&\phantom{\max_{\mbox{\footnotesize$U_\mathrm{A},U_\mathrm{B}$}}\bigl[}
-H_N\bigl(p^\mathrm{(AB)}\bigr)\bigr]
\end{eqnarray}
and then observe that
\begin{eqnarray}
  \label{eq:13}
H_N\bigl(p^\mathrm{(AB)}\bigr)-H_N\bigl(p^\mathrm{(B)}\bigr)
&=&-\!\!\sum_{m,n=1}^N\!\!p^\mathrm{(AB)}_{mn}\log_N
\frac{p^\mathrm{(AB)}_{mn}}{\sum\limits_{m'}p^\mathrm{(AB)}_{m'n}}
\nonumber\\&\geq&0\,.  
\end{eqnarray}
As a consequence, we get 
$V^\mathrm{(AB)}\leq\widetilde{V}^\mathrm{(AB)}\leq 
S_N\bigl(\rho^\mathrm{(A)}\bigr)$.
In conjunction with (\ref{eq:10}) this implies the first inequality in
(\ref{eq:9}), and then the second inequality follows from the symmetry 
of the setup.

Alternatively, we can estimate the two-particle visibility by an
information-theoretic argument. 
When Bob performs his unitary transformation $U_\mathrm{B}$ and then
registers a click of his $n$-th detector, it is \emph{as if} he prepared
Alice's particle in the state described by 
\begin{equation}
\rho^\mathrm{(A)}_n=\frac{1}{p^\mathrm{(B)}_n}
\mathrm{tr}_\mathrm{B}\left\{
P^\mathrm{(B)}_nU_\mathrm{B}^\dagger\rho^\mathrm{(AB)}U_\mathrm{B}\right\}
\,,
\end{equation}
where $P^\mathrm{(B)}_n$ is the projector corresponding to a click of his
$n$-th detector.
In effect, Alice applies her $U_\mathrm{A}$ to $\rho^\mathrm{(A)}_n$
and then detects her particle. 

The process of finding the two-particle visibility in the experiment can now be
understood as a communication protocol where Alice and Bob collaborate 
to obtain the maximal mutual information in the B-to-A quantum channel 
with the restriction that only projective measurements are allowed. 
The well known Holevo bound \cite{Holevo:73}
limits the available mutual information in situations of this kind.
In the particular case under consideration,  
the Holevo bound reads
\begin{equation}
  \label{eq:14}
I\bigl(p^\mathrm{(AB)}\bigr)\leq 
S_N\bigl(\rho^\mathrm{(A)}\bigr)
- \min_{\mbox{\footnotesize$U_\mathrm{B}$}}
\left\{\sum_{n=1}^{N}p^\mathrm{(B)}_nS_N\bigl(\rho^\mathrm{(A)}_n\bigr)\right\}
\,.
\end{equation}
Since the subtracted term is nonnegative, it follows that 
$V^\mathrm{(AB)}\leq S_N\bigl(\rho^\mathrm{(A)}\bigr)$, and the argument
concludes as above.

To prove the second assertion at (\ref{eq:9}), namely that 
the upper bound for the sums of single-particle and two-particle
visibilities in (\ref{eq:9}) is tight, 
we consider the following one-parameter family of pure states 
\begin{equation}
  \label{eq:15}
  \rho^\mathrm{(AB)}_\lambda=|\lambda\rangle\langle\lambda|
\,,\qquad
|\lambda\rangle=\frac{|0\rangle(1-\lambda)+|1\rangle\lambda\sqrt{N}}
{\sqrt{1+(N-1)\lambda^2}}\,,
\end{equation}
where we have, for $\lambda=0$, the ket vector 
\begin{equation}\label{eq:16}
  |0\rangle=\frac{1}{\sqrt{N}}\sum_{k=1}^N|kk\rangle
\end{equation}
of the maximally entangled state (\ref{eq:6}) and, for $\lambda=1$, 
the ket vector 
\begin{equation}
  \label{eq:17}
  |1\rangle=\frac{1}{N}\sum_{j,k=1}^N|jk\rangle
=\bigl(F_\mathrm{A}F_\mathrm{B}\bigr)^{-1}|NN\rangle
\end{equation}
of the product state that is obtained from $|NN\rangle$ by
two-fold inverse Fourier transformation. 
The normalizing denominator in (\ref{eq:15}) takes
$\langle0|1\rangle=1/\sqrt{N}$ into account.

If the source emits the two-particle state $\rho^\mathrm{(AB)}_\lambda$
and all phases $\phi^\mathrm{(A)}_k$,  $\phi^\mathrm{(B)}_l$ are set to zero,
then  the coincidence probabilities are 
$p^\mathrm{(AB)}_{mn}=\bigl|\langle mn|\lambda\rangle\bigr|^2$.
The ingredients of the probability amplitude are
\begin{eqnarray}
  \label{eq:18}
  \langle mn|F_\mathrm{A}F_\mathrm{B}|0\rangle
&=&\frac{\delta_{m+n,N}^{(N)}}{\sqrt{N}}\,,\quad
\nonumber\\
\langle mn|F_\mathrm{A}F_\mathrm{B}|1\rangle&=&\langle mn|NN\rangle
=\delta_{m,N}^{(N)}\delta_{n,N}^{(N)}\,,
\end{eqnarray}
where the periodic Kronecker delta $\delta_{j,k}^{(N)}$ is $1$ if 
$j=k$ modulo $N$, and $0$ otherwise.
The resulting probabilities
\begin{equation}
  \label{eq:19}
p^\mathrm{(AB)}_{mn}
=\frac{\left[(1-\lambda)\delta^{(N)}_{m+n,N}
+N\lambda\delta^{(N)}_{m,N}\delta^{(N)}_{n,N}\right]^2}
{N[1+(N-1)\lambda^2]}
\end{equation}
exhibit perfect correlations inasmuch as for each $m$ value there is only
one $n$ value for which $p^\mathrm{(AB)}_{mn}\neq0$.
As a consequence then, all summands vanish in the double sum of
(\ref{eq:13}), which in turn implies 
\begin{equation}
  \label{eq:20}
  1=\bigl[1-H_N\bigl(p^\mathrm{(A)}\bigr)\bigr]
+ I\bigl(p^\mathrm{(AB)}\bigr)
\leq
V^\mathrm{(A)}+ V^\mathrm{(AB)}\leq1\,,
\end{equation}
so that equal signs must hold throughout. 
Indeed, the upper bound in (\ref{eq:9}) is reached for all $\lambda$, with
$V^\mathrm{(AB)}=1$ for $\lambda=0$ and $V^\mathrm{(AB)}=0$ for $\lambda=1$,
and all intermediate values in between.

\begin{acknowledgments}
M\.Z and BGE gratefully acknowledge the splendid hospitality extended to them
at the National University of Singapore.
This work was supported by A$^*$Star Grant No.\ 012-104-0040,
by KBN project No.\ \mbox{5 PO3B 088 20},
and also by the Telecommunication and Informatics
Task Force (TITF) initiative of Texas A\&M University.  
\end{acknowledgments}

\end{document}